\documentclass[aps,prd,superscriptaddress,twoside,twocolumn,nofootinbib,showpacs,floatfix]{revtex4-1}
\usepackage{amsmath,amssymb}
\usepackage{graphicx}
\usepackage{subfigure}
\pdfpagewidth=\paperwidth
\pdfpageheight=\paperheight
\allowdisplaybreaks
\newcommand*{\slashed}[1]{{#1\!\!\!/}}
\newcommand*{\hc}{\text{H.\,c.}}
\newcommand{\RNum}[1]{\uppercase\expandafter{\romannumeral #1\relax}}
\usepackage{color}

\begin{document}

\title{\boldmath Photoproduction $\gamma p \to K^+\Lambda(1520)$ in an effective Lagrangian approach}

\author{Neng-Chang Wei}
\affiliation{School of Nuclear Science and Technology, University of Chinese Academy of Sciences, Beijing 101408, China}
%\affiliation{School of Physics and Microelectronics, Zhengzhou University, Zhengzhou, Henan 450001, China}

\author{Yu Zhang}
\affiliation{School of Nuclear Science and Technology, University of Chinese Academy of Sciences, Beijing 101408, China}

\author{Fei Huang}
\email{huangfei@ucas.ac.cn}
\affiliation{School of Nuclear Science and Technology, University of Chinese Academy of Sciences, Beijing 101408, China}

\author{De-Min Li}
\email{lidm@zzu.edu.cn}
\affiliation{School of Physics and Microelectronics, Zhengzhou University, Zhengzhou, Henan 450001, China}

\date{\today}

\begin{abstract}
The data on differential cross sections and photon-beam asymmetries for the $\gamma p \to K^+\Lambda(1520)$ reaction have been analyzed within a tree-level effective Lagrangian approach. In addition to the $t$-channel $K$ and $K^\ast$ exchanges, the $u$-channel $\Lambda$ exchange, the $s$-channel nucleon exchange, and the interaction current, a minimal number of nucleon resonances in the $s$ channel are introduced in constructing the reaction amplitudes to describe the data. The results show that the experimental data can be well reproduced by including either the $N(2060)5/2^-$ or the $N(2120)3/2^-$ resonance. In both cases, the contact term and the $K$ exchange are found to make significant contributions, while the contributions from the $K^\ast$ and $\Lambda$ exchanges are negligible in the former case and considerable in the latter case. Measurements of the data on target asymmetries are called on to further pin down the resonance contents and to clarify the roles of the $K^\ast$ and $\Lambda$ exchanges in this reaction.
\end{abstract}



\maketitle

\section{Introduction}   \label{Sec:intro}

The traditional $\pi N$  elastic and inelastic scattering experiments have provided us with abundant knowledge of the mass spectrum and decay properties of the nucleon resonances ($N^\ast$'s). Nevertheless, both the quark model \cite{Isgur:1977ef,Koniuk:1979vy} and lattice QCD \cite{Edwards:2011jj,Edwards:2012fx} calculations predict more resonances than have been observed in the $\pi N$ scattering experiments. The resonances predicated by quark model or lattice QCD but not observed in experiments are called ``missing resonances'', which are supposed to have small couplings to the $\pi N$ channel and, thus, escape from experimental detection. In the past few decades, intense efforts have been dedicated to search for the missing resonances in meson production reaction channels other than $\pi N$. In particular, the $\rho N$, $\phi N$, and $\omega N$ production reactions in the nonstrangeness sector and the $KY$, $K^\ast Y$ ($Y=\Lambda, \Sigma$) production reactions in the strangeness sector have been widely investigated both experimentally and theoretically. 

In the present paper, we focus on the $\gamma p \to K^+\Lambda(1520)$ reaction process. The threshold of the $K^+\Lambda(1520)$ photoproduction is about $2.01$ GeV, and, thus, this reaction provides a chance to study the $N^\ast$ resonances in the $W\sim 2.0$ GeV mass region in which we have infancy information as shown in the latest version of the Review of Particle Physics (RPP) \cite{Tanabashi:2018oca}. Besides, the isoscalar nature of $\Lambda(1520)$ allows only the $I=1/2$ $N^\ast$ resonances exchanges in the $s$ channel, which simplifies the reaction mechanisms of the $K^+\Lambda(1520)$ photoproduction.

Experimentally, the cross sections for the reaction $\gamma p \to K^+\Lambda(1520)$ have been measured at SLAC by Boyarski {\it et al.} in 1971 for photon energy $E_\gamma=11$ GeV \cite{Boyarski:1970yc}, and by the LAMP2 group in 1980 at $E_\gamma=2.8$$-$$4.8$ GeV \cite{Barber:1980zv}. In 2010, the LEPS Collaboration measured the differential cross sections and photon-beam asymmetries ($\Sigma$) at Spring-8 for $\gamma p \to K^+\Lambda(1520)$ at energies from threshold up to $E_\gamma=2.6$ GeV at forward $K^+$ angles \cite{Kohri:2009xe}. In 2011, the SAPHIR Collaboration measured the cross sections at the Electron Stretcher Accelerator (ELSA) for the $K^+\Lambda(1520)$ photoproduction in the energy range from threshold up to $E_\gamma=2.65$ GeV \cite{Wieland:2010cq}. Recently, the differential and total cross sections for the $K^+\Lambda(1520)$ photoproduction were reported by the CLAS Collaboration at energies from threshold up to the center-of-mass energy $W=2.86$ GeV over a large range of the $K^+$ production angle \cite{Moriya:2013hwg}.

Theoretically, the $K^+\Lambda(1520)$ photoproduction reaction has been extensively investigated based on effective Lagrangian approaches by four theory groups in $11$ publications \cite{Nam:2005uq,Nam:2006cx,Nam:2009cv,Nam:2010au,Toki:2007ab,Xie:2010yk,Xie:2013mua,Wang:2014jxb,He:2012ud,He:2014gga,Yu:2017kng}. In Refs.~\cite{Nam:2005uq,Nam:2006cx,Nam:2009cv,Nam:2010au}, Nam {\it et al.} found  that the contact term and the $t$-channel $K$ exchange are important to the cross sections of $\gamma p \to K^+\Lambda(1520)$, while the contributions from the $t$-channel $K^\ast$ exchange and the $s$-channel nucleon resonance exchange are rather small. In Refs.~\cite{Toki:2007ab,Xie:2010yk,Xie:2013mua,Wang:2014jxb}, Xie, Wang, and Nieves {\it et al.} found that apart from the contact term and the $t$-channel $K$ exchange, the $u$-channel $\Lambda$ exchange and the $s$-channel $N(2120)3/2^-$ [previously called $D_{13}(2080)$] exchange are also important in describing the cross-section data for $\gamma p \to K^+\Lambda(1520)$, while the contribution from the $t$-channel $K^\ast$ exchange is negligible in this reaction. In Refs.~\cite{He:2012ud,He:2014gga}, He and Chen found that the contribution from the $t$-channel $K^\ast$ exchange in $\gamma p \to K^+\Lambda(1520)$ is also considerable besides the important contributions from the contact term, the $t$-channel $K$ exchange, the $u$-channel $\Lambda$ exchange, and the $s$-channel $N(2120)3/2^-$ exchange. In Ref.~\cite{Yu:2017kng}, Yu and Kong studied the $\gamma p \to K^+\Lambda(1520)$ reaction within a Reggeized model, and they claimed that the important contributions to this reaction are coming from the contact term, the $t$-channel $K$ exchange, and the $t$-channel $K^\ast_2$ exchange, while the contribution from the $t$-channel $K^\ast$ exchange is minor. 

One observes that the common feature reported in all the above-mentioned publications of Refs.~\cite{Nam:2005uq,Nam:2006cx,Nam:2009cv,Nam:2010au,Toki:2007ab,Xie:2010yk,Xie:2013mua,Wang:2014jxb,He:2012ud,He:2014gga,Yu:2017kng} is that the contributions from the contact term and the $t$-channel $K$ exchange are important to the $\gamma p \to K^+\Lambda(1520)$ reaction. Even so, the reaction mechanisms of $\gamma p \to K^+\Lambda(1520)$ claimed by those four theory groups are quite different. In particular, there are no conclusive answers which can be derived from Refs.~\cite{Nam:2005uq,Nam:2006cx,Nam:2009cv,Nam:2010au,Toki:2007ab,Xie:2010yk,Xie:2013mua,Wang:2014jxb,He:2012ud,He:2014gga,Yu:2017kng} for the following questions: Are the contributions from the $t$-channel $K^\ast$ exchange and $u$-channel $\Lambda$ exchange significant or not in this reaction, does one inevitably need to introduce nucleon resonances in the $s$ channel to describe the data, and if yes, is the $N(2120)3/2^-$ resonance the only candidate needed in this reaction and what are the parameters of it?

\begin{figure}[tbp]
\includegraphics[width=0.9\columnwidth]{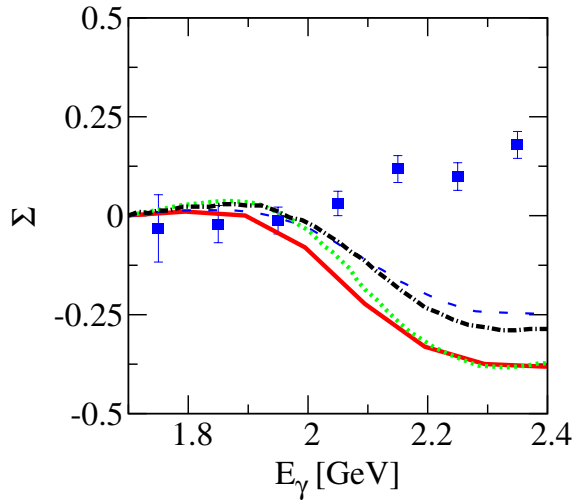}
\caption{Predictions of photon-beam asymmetries at $\cos\theta=0.8$ as a function of the photon laboratory energy for $\gamma p \to K^+\Lambda(1520)$ from Ref.~\cite{Xie:2010yk} (blue dashed line), the fit \RNum{2} of Ref.~\cite{Xie:2013mua} (red solid line), Ref.~\cite{He:2012ud} (green dotted line), and Ref.~\cite{He:2014gga} (black dot-dashed line). The data are located in $0.6<\cos\theta<1$ and taken from the LEPS Collaboration \cite{Kohri:2009xe} (blue square).}
\label{fit:bema-xie}
\end{figure}

On the other hand, the data on photon-beam asymmetries for $\gamma p \to K^+\Lambda(1520)$ reported by the LEPS Collaboration in 2010 \cite{Kohri:2009xe} have never been well reproduced in previous publications of Refs.~\cite{Nam:2005uq,Nam:2006cx,Nam:2009cv,Nam:2010au,Toki:2007ab,Xie:2010yk,Xie:2013mua,Wang:2014jxb,He:2012ud,He:2014gga}. As an illustration, we show in Fig.~\ref{fit:bema-xie} the theoretical results on photon-beam asymmetries from Refs.~\cite{Xie:2010yk,He:2012ud,Xie:2013mua,He:2014gga} calculated at $\cos\theta=0.8$ and compared with the data located at $0.6<\cos\theta<1$. It is true that the data bins in scattering angles are wide; nevertheless, it has been checked that the averaged values of theoretical beam-asymmetry results in $0.6<\cos\theta<1$ are comparable with those calculated at $\cos\theta=0.8$. One sees that, in the energy region $E_\gamma > 2$ GeV, even the signs of the photon-beam asymmetries predicated by these theoretical works are opposite to the data. In the Regge model analysis of Ref.~\cite{Yu:2017kng}, the photon-beam asymmetries have indeed been analyzed, but there the differential cross-section data have been only qualitatively described, and the structures of the angular distributions exhibited by the data were missing due to the lack of nucleon resonances in the $s$-channel interactions.

The purpose of the present work is to perform a combined analysis of the available data on both the differential cross sections and the photon-beam asymmetries for $\gamma p \to K^+\Lambda(1520)$ within an effective Lagrangian approach, and, based on that, we try to get a clear understanding of the reaction mechanism of $\gamma p \to K^+\Lambda(1520)$. In particular, we aim to clarify whether the $t$-channel $K^\ast$ exchange and the $u$-channel $\Lambda$ exchange are important or not and what the resonance contents and their associated parameters are in this reaction. As discussed above, previous publications of Refs.~\cite{Nam:2005uq,Nam:2006cx,Nam:2009cv,Nam:2010au,Toki:2007ab,Xie:2010yk,Xie:2013mua,Wang:2014jxb,He:2012ud,He:2014gga} can describe only the differential cross-section data, and they gave diverse answers to these questions. It is expected that more reliable results on the resonance contents and the roles of $K^\ast$ and $\Lambda$ exchanges in this reaction can be obtained from the theoretical analysis which can result in a satisfactory description of the data on both the differential cross sections and the photon-beam asymmetries.  

The present paper is organized as follows. In Sec.~\ref{Sec:formalism}, we briefly introduce the framework of our theoretical model, including the effective interaction Lagrangians, the resonance propagators, and the phenomenological form factors employed in this work. The results of our model calculations are shown and discussed in Sec.~\ref{Sec:results}. Finally, a brief summary and conclusions are given in Sec.~\ref{sec:summary}.

\section{Formalism}  \label{Sec:formalism}

\begin{figure}[tbp]
\centering
{\vglue 0.15cm}
\subfigure[~$s$ channel]{
\includegraphics[width=0.45\columnwidth]{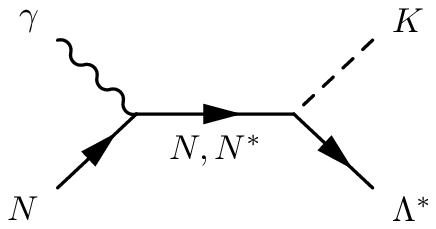}}  {\hglue 0.4cm}
\subfigure[~$t$ channel]{
\includegraphics[width=0.45\columnwidth]{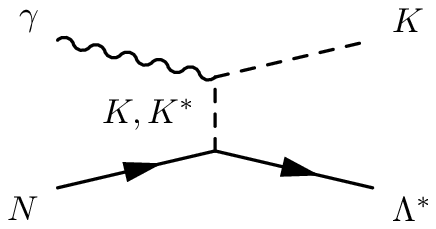}} \\[6pt]
\subfigure[~$u$ channel]{
\includegraphics[width=0.45\columnwidth]{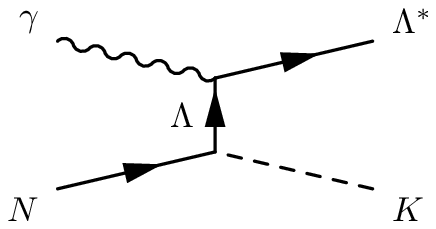}} {\hglue 0.4cm}
\subfigure[~Interaction current]{
\includegraphics[width=0.45\columnwidth]{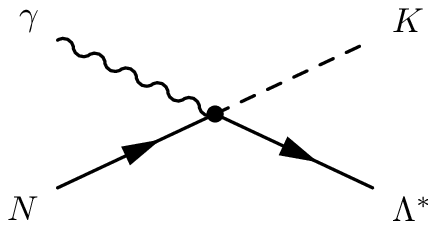}}
\caption{Generic structure of the amplitude for $\gamma p \to K^+\Lambda(1520)$. Time proceeds from left to right. The outgoing $\Lambda^\ast$ denotes $\Lambda(1520)$.}
\label{FIG:feymans}
\end{figure}

The full amputated photoproduction amplitude for $\gamma N \to K\Lambda(1520)$ in our tree-level effective Lagrangian approach can be expressed as \cite{Wang:2017tpe,Wang:2018vlv,Wang:2020mdn,Wei:2019imo}
\begin{equation}
M^{\nu\mu} \equiv M^{\nu\mu}_s + M^{\nu\mu}_t + M^{\nu\mu}_u + M^{\nu\mu}_{\rm int},  \label{eq:amplitude}
\end{equation}
with $\nu$ and $\mu$ being the Lorentz indices for outgoing $\Lambda(1520)$ and incoming photon, respectively. The first three terms $M^{\nu\mu}_s$, $M^{\nu\mu}_t$, and $M^{\nu\mu}_u$ stand for the amplitudes resulted from the $s$-channel $N$ and $N^\ast$ exchanges, the $t$-channel $K$ and $K^\ast$ exchanges, and the $u$-channel $\Lambda$ exchange, respectively, as diagrammatically depicted in Fig.~\ref{FIG:feymans}. They can be calculated straightforward by using the effective Lagrangians, propagators, and form factors provided in the following part of this section. The last term in Eq.~(\ref{eq:amplitude}) represents the interaction current arising from the photon attaching to the internal structure of the $\Lambda(1520)NK$ vertex. In practical calculation, the interaction current $M^{\nu\mu}_{\rm int}$ is modeled by a generalized contact current \cite{Haberzettl:1997,Haberzettl:2006,Haberzettl:2011zr,Huang:2012,Huang:2013,Wang:2017tpe,Wang:2018vlv,Wei:2019imo,Wang:2020mdn,Wei:2020fmh}:
\begin{equation}
M^{\nu\mu}_{\rm int} = \Gamma_{\Lambda^\ast NK}^\nu(q) C^\mu + M^{\nu\mu}_{\rm KR} f_t.   \label{eq:Mint}
\end{equation}
Here $\Gamma_{\Lambda^\ast NK}^\nu(q)$ is the vertex function of $\Lambda(1520)NK$ coupling governed by the Lagrangian of Eq.~(\ref{eq:lpls}):
\begin{equation}
\Gamma_{\Lambda^\ast NK}^\nu(q) = - \frac{g_{\Lambda^\ast NK}}{M_K} \gamma_5 q^\nu,
\end{equation}
with $q$ being the four-momentum of the outgoing $K$ meson; $M^{\nu\mu}_{\rm KR}$ is the Kroll-Ruderman term governed by the Lagrangian of Eq.~(\ref{eq:con}):
\begin{equation}
M^{\nu\mu}_{\rm KR} = \frac{g_{\Lambda^\ast NK}}{M_K} g^{\nu\mu} \gamma_5 Q_K \tau,
\end{equation}
with $Q_K$ being the electric charge of outgoing $K$ meson and $\tau$ being the isospin factor of the Kroll-Ruderman term; $f_t$ is the phenomenological form factor attached to the amplitude of $t$-channel $K$ exchange, which is given by Eq.~(\ref{eq:form_factor_t}); $C^\mu$ is an auxiliary current introduced to ensure the gauge invariance of the full photoproduction amplitude of Eq.~(\ref{eq:amplitude}). Note that the photoproduction amplitudes will automatically be gauge invariant in the cases that there are no form factors and the electromagnetic couplings are obtained by replacing the partial derivative by its covariant form in the corresponding hadronic vertices. In practical calculation, one has to introduce the form factors in hadronic vertices (cf. Sec.~\ref{subSec:form factor}) which violate the gauge invariance. The auxiliary current $C^\mu$ is then introduced to compensate the gauge violation caused by the form factors. Following Refs.~\cite{Haberzettl:2006,Haberzettl:2011zr,Huang:2012}, for the $\gamma N \to K\Lambda(1520)$ reaction, the auxiliary current $C^\mu$ is chosen to be
\begin{equation}
C^\mu =  - Q_K \tau \frac{f_t-\hat{F}}{t-q^2}  (2q-k)^\mu - \tau Q_N \frac{f_s-\hat{F}}{s-p^2} (2p+k)^\mu,  \label{eq:GI-auxi}
\end{equation}
with
\begin{equation} \label{eq:Fhat}
\hat{F} = 1 - \hat{h} \left(1 -  f_s\right) \left(1 - f_t\right).
\end{equation}
Here $p$, $q$, and $k$ denote the four-momenta for incoming $N$, outgoing $K$, and incoming photon, respectively; $Q_K$ and $Q_N$ are electric charges of $K$ and $N$, respectively; $f_s$ and $f_t$ are phenomenological form factors for $s$-channel $N$ exchange and $t$-channel $K$ exchange, respectively; $\hat{h}$ is an arbitrary function going to unity in the high-energy limit and set to be $1$ in the present work for simplicity; $\tau$ depicts the isospin factor for the corresponding hadronic vertex. Alternatively, one can rewrite the auxiliary current $C^\mu$ in Eq.~(\ref{eq:GI-auxi}) as
\begin{align}
C^\mu = & - Q_K \tau (2q-k)^\mu \frac{f_t-1}{t-q^2} \left[1-\hat{h}\left(1-f_s\right) \right]  \nonumber \\
 & - \tau Q_N (2p+k)^\mu \frac{f_s-1}{s-p^2} \left[1-\hat{h} \left(1-f_t\right) \right].
\end{align}
One sees clearly that if there are no form factors, i.e., $f_t = f_s = 1$, one has $C^\mu \to 0$ and, consequently, $M^{\nu\mu}_{\rm int} \to M^{\nu\mu}_{KR}$. We mention that the auxiliary current $C^\mu$ in Eq.~(\ref{eq:GI-auxi}) works for both real and virtual photons; i.e., the amplitudes we constructed in Eq.~(\ref{eq:amplitude}) are gauge invariant for both photo- and electroproduction of $K^+\Lambda(1520)$. In Ref.~\cite{Nam2013}, another prescription for keeping gauge invariance of the $K^+\Lambda(1520)$ electroproduction amplitudes was introduced, where additional terms are considered besides those for photoproduction reactions.

In the rest of this section, we present the effective Lagrangians, the resonance propagators, the form factors, and the interpolated $t$-channel Regge amplitudes employed in the present work.

\subsection{Effective Lagrangians} \label{Sec:Lagrangians}

In this subsection, we list all the Lagrangians used in the present work. For further simplicity, we define the operators
\begin{equation}
\Gamma^{(+)}=\gamma_5 \qquad {\rm and} \qquad \Gamma^{(-)}=1,
\end{equation}
the field
\begin{equation}
\Lambda^\ast = \Lambda(1520),
\end{equation}
and the field-strength tensor
\begin{equation}
F^{\mu\nu} = \partial^\mu A^\nu - \partial^\nu A^\mu, 
\end{equation}
with $A^\mu$ denoting the electromagnetic field.

The Lagrangians needed to calculate the amplitudes for nonresonant interacting diagrams are
\begin{eqnarray}
{\cal L}_{\gamma NN} &=& -\,e \bar{N} \!\left[ \! \left( \hat{e} \gamma^\mu - \frac{ \hat{\kappa}_N} {2M_N}\sigma^{\mu \nu}\partial_\nu \! \right) \! A_\mu\right]\! N, \\[6pt]
{\cal L}_{\gamma KK} &=& ie \!\left[K^+\left(\partial_\mu K^-\right)-K^-\left(\partial_\mu K^+\right)\right] \! A^\mu, \\[6pt]
{\cal L}_{\gamma K{K^\ast}} &=& e\frac{g_{\gamma K{K^\ast}}}{M_K}\varepsilon^{\alpha \mu \lambda \nu}\left(\partial_\alpha A_\mu\right)\left(\partial_\lambda K\right) K^\ast_\nu, \label{Lag:gKKst}   \\[6pt]
{\cal L}_{\gamma \Lambda\Lambda^\ast} &=& - \, ie \frac{g^{(1)}_{\Lambda^\ast\Lambda\gamma}}{2M_\Lambda}{\bar{\Lambda}}^{\ast\mu} \gamma^\nu F_{\mu\nu} \Lambda  \nonumber  \\
               && + \, e \frac{g^{(2)}_{\Lambda^\ast\Lambda\gamma}}{(2M_\Lambda)^2} {\bar{\Lambda}}^{\ast\mu} F_{\mu\nu} \partial^\nu\Lambda  +   \hc,  \label{eq:gll} \\[6pt]
{\cal L}_{\gamma \Lambda^\ast NK} &=& -iQ_K\frac{g_{\Lambda^\ast NK}}{M_K} \bar{\Lambda}^{^\ast \mu} A_\mu K\gamma_5 N + \hc, \label{eq:con}  \\[6pt]
{\cal L}_{\Lambda^\ast NK} &=& \frac{g_{\Lambda^\ast NK}}{M_K} {\bar{\Lambda}}^{\ast \mu} \left(\partial_\mu K\right)\gamma_5 N + \hc, \label{eq:lpls} \\[6pt]
{\cal L}_{\Lambda^\ast N K^\ast} &=& -\frac{ig_{\Lambda^\ast N K^\ast}}{M_{K^\ast}}\bar{\Lambda}^{^\ast \mu}\gamma^\nu \left(\partial_\mu K^{\ast}_\nu - \partial_\nu K^{\ast}_\mu \right)N  \nonumber \\
                                && + \, \hc,  \\[6pt]
{\cal L}_{\Lambda N K} &=& -ig_{\Lambda NK}\bar{\Lambda}\gamma_5 K N + \hc,
\end{eqnarray}
where $M_{K^\ast}$, $M_K$, $M_N$, and $M_\Lambda$ denote the masses of $K^\ast$, $K$, $N$, and $\Lambda$, respectively; $\hat{e}$ stands for the charge operator and $\hat{\kappa}_N = \kappa_p\left(1+\tau_3\right)/2 + \kappa_n\left(1-\tau_3\right)/2$ with the anomalous magnetic moments $\kappa_p=1.793$ and $\kappa_n=-1.913$. The coupling constant $g_{\gamma K K^{\ast}} = 0.413$ is calculated by the radiative decay width of $K^\ast\to K\gamma$ given by RPP \cite{Tanabashi:2018oca} with the sign inferred from $g_{\gamma \pi \rho}$ \cite{Garcilazo:1993av} via the flavor SU(3) symmetry considerations in conjunction with the vector-meson dominance assumption. The coupling constants $g^{(1)}_{\Lambda^\ast\Lambda\gamma}$ and $g^{(2)}_{\Lambda^\ast\Lambda\gamma}$ are fit parameters, but only one of them is free since they are constrained by the $\Lambda(1520)$ radiative decay width $\Gamma_{\Lambda(1520) \to \Lambda\gamma}=0.133$ MeV as given by RPP \cite{Tanabashi:2018oca}. The value of $g_{\Lambda^\ast NK} =10.5$ is determined by the decay width of $\Lambda(1520) \to NK$, $\Gamma_{\Lambda(1520) \to NK}=7.079$ MeV, as advocated by RPP \cite{Tanabashi:2018oca}. The coupling constant $g_{\Lambda^\ast N K^\ast}$ is a parameter to be determined by fitting the data. The coupling constant $g_{\Lambda NK}\approx-14$ is determined by the flavor SU(3) symmetry, $g_{\Lambda NK} = \left(-3\sqrt{3} /5\right) g_{NN\pi}$ with $g_{NN\pi}=13.46$.  

For nucleon resonances in the $s$ channel, the Lagrangians for electromagnetic couplings read \cite{Wang:2017tpe,Wang:2018vlv,Wei:2019imo,Wang:2020mdn}
\begin{eqnarray}
{\cal L}_{RN\gamma}^{1/2\pm} &=& e\frac{g_{RN\gamma}^{(1)}}{2M_N}\bar{R} \Gamma^{(\mp)}\sigma_{\mu\nu} \left(\partial^\nu A^\mu \right) N  + \hc, \\[6pt]
{\cal L}_{RN\gamma}^{3/2\pm} &=& -\, ie\frac{g_{RN\gamma}^{(1)}}{2M_N}\bar{R}_\mu \gamma_\nu \Gamma^{(\pm)}F^{\mu\nu}N \nonumber \\
&&+\, e\frac{g_{RN\gamma}^{(2)}}{\left(2M_N\right)^2}\bar{R}_\mu \Gamma^{(\pm)}F^{\mu \nu}\partial_\nu N + \hc, \\[6pt]
{\cal L}_{RN\gamma}^{5/2\pm} & = & e\frac{g_{RN\gamma}^{(1)}}{\left(2M_N\right)^2}\bar{R}_{\mu \alpha}\gamma_\nu \Gamma^{(\mp)}\left(\partial^{\alpha} F^{\mu \nu}\right)N \nonumber \\
&& \pm\, ie\frac{g_{RN\gamma}^{(2)}}{\left(2M_N\right)^3}\bar{R}_{\mu \alpha} \Gamma^{(\mp)}\left(\partial^\alpha F^{\mu \nu}\right)\partial_\nu N \nonumber \\
&& + \,  \hc,  \\[6pt]
{\cal L}_{RN\gamma}^{7/2\pm} &=&  ie\frac{g_{RN\gamma}^{(1)}}{\left(2M_N\right)^3}\bar{R}_{\mu \alpha \beta}\gamma_\nu \Gamma^{(\pm)}\left(\partial^{\alpha}\partial^{\beta} F^{\mu \nu}\right)N \nonumber \\
&&-\, e\frac{g_{RN\gamma}^{(2)}}{\left(2M_N\right)^4}\bar{R}_{\mu \alpha \beta} \Gamma^{(\pm)} \left(\partial^\alpha \partial^\beta F^{\mu \nu}\right) \partial_\nu N  \nonumber \\
&&  + \,  \hc,
\end{eqnarray}
and the Lagrangians for hadronic couplings to $\Lambda(1520)K$ read
\begin{eqnarray}
{\cal L}_{R\Lambda^\ast K}^{1/2\pm} &=& \frac{g^{(1)}_{R\Lambda^\ast K}}{M_K} \bar{\Lambda}^{\ast\mu}\Gamma^{(\pm)} \left(\partial_\mu K\right) R + \hc,  \\[6pt]
{\cal L}_{R\Lambda^\ast K}^{3/2\pm} &=& \frac{g^{(1)}_{R\Lambda^\ast K}}{M_K} \bar{\Lambda}^{\ast\mu}\gamma_\nu \Gamma^{(\mp)} \left(\partial^\nu K\right) R_\mu \nonumber \\
&& + \, i \frac{g^{(2)}_{R\Lambda^\ast K}}{M_K^2} \bar{\Lambda}^\ast_\alpha \Gamma^{(\mp)} \left(\partial^\mu \partial^\alpha K\right) R_\mu  + \hc, \\[6pt]
{\cal L}_{R\Lambda^\ast K}^{5/2\pm} &=& i\frac{g^{(1)}_{R\Lambda^\ast K}}{M_K^2}  \bar{\Lambda}^{\ast\alpha}\gamma_\mu \Gamma^{(\pm)} \left(\partial^\mu \partial^\beta K\right) R_{\alpha\beta} \nonumber  \\
&&- \,\frac{g^{(2)}_{R\Lambda^\ast K}}{M_K^3} \bar{\Lambda}^\ast_\mu \Gamma^{(\pm)} \left(\partial^\mu \partial^\alpha \partial^\beta K\right) R_{\alpha\beta} \nonumber \\
&& + \, \hc, \\[6pt]
{\cal L}_{R\Lambda^\ast K}^{7/2\pm} &=& -\frac{g^{(1)}_{R\Lambda^\ast K}}{M_K^3} \bar{\Lambda}^{\ast\alpha}\gamma_\mu \Gamma^{(\mp)} \left(\partial^\mu \partial^\beta \partial^\lambda K\right) R_{\alpha\beta\lambda} \nonumber \\
&& - \, i \frac{g^{(2)}_{R\Lambda^\ast K}}{M_K^4} \bar{\Lambda}^{\ast}_\mu \Gamma^{(\mp)} \left(\partial^\mu \partial^\alpha \partial^\beta \partial^\lambda K\right) R_{\alpha\beta\lambda} \nonumber \\
&& + \, \hc,
\end{eqnarray}
where $R$ designates the $N^\ast$ resonance and the superscript of ${\cal L}_{RN\gamma}$ and ${\cal L}_{R\Lambda^\ast K}$ denotes the spin and parity of the resonance $R$. The coupling constants $g_{RN\gamma}^{(i)}$ and $g^{(i)}_{R\Lambda^\ast K}$ $(i=1,2)$ are fit parameters. Actually, only the products of $g_{RN\gamma}^{(i)} g^{(j)}_{R\Lambda^\ast K}$ $(i,j=1,2)$ are relevant to the reaction amplitudes, and they are what we really fit in practice.

In Ref.~\cite{Nam:2005uq}, the off-shell effects for spin-$3/2$ resonances in $\gamma p\to K^+ \Lambda(1520)$ have been tested. It was found that the off-shell effects are small and the off-shell parameter $X$ can be set to zero. In the present work, we simply ignore the off-shell terms in the interaction Lagrangians for high spin resonances and leave this issue for future work.

%In Refs.~\cite{Benmerrouche89,Davidson91,Mizutani98,Mizutani99}, the off-shell effects of spin-$3/2$ resonances have been investigated in pion and kaon electromagnetic production reactions. In the above-mentioned Lagrangians for resonance couplings, the off-shell effects have been simply ignored. We leave this issue for future work.

\subsection{Resonance propagators}

We follow Ref.~\cite{Wang:2017tpe} to use the following prescriptions for the propagators of resonances with spin $1/2$, $3/2$, $5/2$, and $7/2$:
\begin{eqnarray}
S_{1/2}(p) &=& \frac{i}{\slashed{p} - M_R + i \Gamma_R/2}, \label{propagator-1hf}  \\[6pt]
S_{3/2}(p) &=&  \frac{i}{\slashed{p} - M_R + i \Gamma_R/2} \left( \tilde{g}_{\mu \nu} + \frac{1}{3} \tilde{\gamma}_\mu \tilde{\gamma}_\nu \right),  \\[6pt]
S_{5/2}(p) &=&  \frac{i}{\slashed{p} - M_R + i \Gamma_R/2} \,\bigg[ \, \frac{1}{2} \big(\tilde{g}_{\mu \alpha} \tilde{g}_{\nu \beta} + \tilde{g}_{\mu \beta} \tilde{g}_{\nu \alpha} \big)  \nonumber \\
&& -\, \frac{1}{5}\tilde{g}_{\mu \nu}\tilde{g}_{\alpha \beta}  + \frac{1}{10} \big(\tilde{g}_{\mu \alpha}\tilde{\gamma}_{\nu} \tilde{\gamma}_{\beta} + \tilde{g}_{\mu \beta}\tilde{\gamma}_{\nu} \tilde{\gamma}_{\alpha}  \nonumber \\
&& +\, \tilde{g}_{\nu \alpha}\tilde{\gamma}_{\mu} \tilde{\gamma}_{\beta} +\tilde{g}_{\nu \beta}\tilde{\gamma}_{\mu} \tilde{\gamma}_{\alpha} \big) \bigg], \\[6pt]
S_{7/2}(p) &=&  \frac{i}{\slashed{p} - M_R + i \Gamma_R/2} \, \frac{1}{36}\sum_{P_{\mu} P_{\nu}} \bigg( \tilde{g}_{\mu_1 \nu_1}\tilde{g}_{\mu_2 \nu_2}\tilde{g}_{\mu_3 \nu_3} \nonumber \\
&& -\, \frac{3}{7}\tilde{g}_{\mu_1 \mu_2}\tilde{g}_{\nu_1 \nu_2}\tilde{g}_{\mu_3 \nu_3} + \frac{3}{7}\tilde{\gamma}_{\mu_1} \tilde{\gamma}_{\nu_1} \tilde{g}_{\mu_2 \nu_2}\tilde{g}_{\mu_3 \nu_3} \nonumber \\
&& -\, \frac{3}{35}\tilde{\gamma}_{\mu_1} \tilde{\gamma}_{\nu_1} \tilde{g}_{\mu_2 \mu_3}\tilde{g}_{\nu_2 \nu_3} \bigg),  \label{propagator-7hf}
\end{eqnarray}
where
\begin{eqnarray}
\tilde{g}_{\mu \nu} &=& -\, g_{\mu \nu} + \frac{p_{\mu} p_{\nu}}{M_R^2}, \\[6pt]
\tilde{\gamma}_{\mu} &=& \gamma^{\nu} \tilde{g}_{\nu \mu} = -\gamma_{\mu} + \frac{p_{\mu}\slashed{p}}{M_R^2},   \label{eq:prop-auxi}
\end{eqnarray}
and the summation over $P_\mu$ $\left(P_\nu\right)$ in Eq.~(\ref{propagator-7hf}) goes over the $3!=6$ possible permutations of the indices $\mu_1\mu_2\mu_3$ $\left(\nu_1\nu_2\nu_3\right)$. In Eqs.~(\ref{propagator-1hf})$-$(\ref{eq:prop-auxi}), $M_R$ and $\Gamma_R$ are the mass and width of resonance $R$ with four-momentum $p$, respectively.

\subsection{Form factors} \label{subSec:form factor}

In practical calculation of the reaction amplitudes, a phenomenological form factor is introduced in each hadronic vertex. For the $t$-channel meson exchanges, we adopt the following form factor  \cite{Wang:2017tpe,Wang:2020mdn,Wang:2018vlv,Wei:2019imo}:
\begin{eqnarray}
f_M(q^2_M) =  \left (\frac{\Lambda_M^2-M_M^2}{\Lambda_M^2-q^2_M} \right)^2,   \label{eq:ff_M}
\end{eqnarray}
and for the $s$-channel and $u$-channel baryon exchanges, we use \cite{Wang:2017tpe,Wang:2020mdn,Wang:2018vlv,Wei:2019imo}
\begin{eqnarray}
f_B(p^2_x) =\left (\frac{\Lambda_B^4}{\Lambda_B^4+\left(p_x^2-M_B^2\right)^2} \right )^2.  \label{eq:ff_B}
\end{eqnarray}
Here, $q_M$ denotes the four-momentum of the intermediate meson in the $t$ channel, and $p_x$ stands for the four-momentum of the intermediate baryon in $s$ and $u$ channels with $x=$ $s$ and $u$, respectively. $\Lambda_{M(B)}$ is the corresponding cutoff parameter. In the present work, in order to reduce the number of adjustable parameters, we use the same cutoff parameter $\Lambda_B$ for all the nonresonant diagrams, i.e., $\Lambda_B \equiv \Lambda_K=\Lambda_{K^\ast}=\Lambda_\Lambda=\Lambda_N$. The parameter $\Lambda_B$ and the cutoff parameter $\Lambda_R$ for $N^\ast$ resonances are determined by fitting the experimental data.

\subsection{Interpolated $t$-channel Regge amplitudes}

A Reggeized treatment of the $t$-channel $K$ and $K^\ast$ exchanges is usually employed to economically describe the high-energy data, which corresponds to the following replacement of the form factors in Feynman amplitudes:
\begin{align}
f_K(q_K^2) \to {\cal F}_K(q_K^2) = & \left(\frac{s}{s_0}\right)^{\alpha_K(t)} \frac{\pi\alpha^\prime_K } {\sin[\pi\alpha_K(t)] }  \nonumber \\
& \times \frac{t-M^2_K} {\Gamma[1+\alpha_K(t)]}, \label{eq:Regge_prop_K} \\[6pt]
f_{K^\ast}(q_{K^\ast}^2) \to {\cal F}_{K^\ast}(q_{K^\ast}^2) = & \left(\frac{s}{s_0}\right)^{\alpha_{K^\ast}(t)-1} \frac{\pi\alpha^\prime_{K^\ast}} {\sin[\pi\alpha_{K^\ast}(t)]}  \nonumber  \\
& \times \frac{t-M^2_{K^\ast}} {\Gamma[\alpha_{K^\ast}(t)]}.  \label{eq:Regge_prop_Kstar}
\end{align}
Here $s_0$ is a mass scale which is conventionally taken as $s_0=1$ GeV$^2$, and $\alpha'_M$ is the slope of the Regge trajectory $\alpha_M(t)$. For $M=K$ and $K^\ast$, the trajectories are parameterized as \cite{Wang:2019mid}
\begin{align}
\alpha_K(t) &= 0.7~{\rm GeV}^{-2}\left(t-m_K^2\right),   \label{eq:trajectory_K}  \\[6pt]
\alpha_{K^\ast}(t) &= 1 + 0.85~{\rm GeV}^{-2} \left(t-m_{K^\ast}^2\right).  \label{eq:trajectory_Kstar}
\end{align}
Note that, in Eqs.~(\ref{eq:Regge_prop_K}) and (\ref{eq:Regge_prop_Kstar}), degenerate trajectories are employed for $K$ and $K^\ast$ exchanges; thus, the signature factors reduce to $1$.

In the present work, we use the so-called interpolated Regge amplitudes for the $t$-channel $K$ and $K^\ast$ exchanges. The idea of this prescription is that at high energies and small angles one uses Regge amplitudes, and at low energies one uses Feynman amplitudes, while in the intermediate energy region an interpolating form factor is introduced to ensure a smooth transition from the low-energy Feynman amplitudes to the high-energy Regge amplitudes. This hybrid Regge approach has been applied to study the $\gamma p \to K^+\Lambda(1520)$ reaction in Refs.~\cite{Nam:2010au,Wang:2014jxb,He:2014gga,Yu:2017kng} and the other reactions in Refs.~\cite{Wang:2015hfm,Wang:2017plf,Wang:2017qcw,Wang:2019mid}. Instead of making the replacements of Eqs.~(\ref{eq:Regge_prop_K}) and (\ref{eq:Regge_prop_Kstar}) in a pure Reggeized treatment, in this hybrid Regge model the amplitudes for $t$-channel $K$ and $K^\ast$ exchanges are constructed by making the following replacements of the form factors in the corresponding Feynman amplitudes:
\begin{equation}
 f_M(q_M^2) \to  {\cal F}_{R,M} = {\cal F}_M(q_M^2)R + f_M(q_M^2) \left(1-R\right),   \label{eq:form_factor_t}
\end{equation}
where ${\cal F}_M(q_M^2)$ $(M=K, K^\ast)$ is defined in Eqs.~(\ref{eq:Regge_prop_K}) and (\ref{eq:Regge_prop_Kstar}) and $R=R_sR_t$ with
\begin{align}
R_s = & \frac{1}{1+e^{-(s-s_R)/s_0}},  \nonumber \\[6pt]
R_t = & \frac{1}{1+e^{-(t+t_R)/t_0}}.
\end{align}
Here $s_R$, $t_R$, $s_0$, and $t_0$ are parameters to be determined by fitting the experimental data.

The auxiliary current $C^\mu$ introduced in Eq.~(\ref{eq:GI-auxi}) and the interaction current $M^{\nu\mu}_{\rm int}$ given in Eq.~(\ref{eq:Mint}) ensures that the full photoproduction amplitude of Eq.~(\ref{eq:amplitude}) satisfies the generalized Ward-Takahashi identity and, thus, is fully gauge invariant \cite{Haberzettl:2006,Haberzettl:2011zr,Huang:2012}. Note that our prescription for $C^\mu$ and $M^{\nu\mu}_{\rm int}$ is independent of any particular form of the $t$-channel form factor $f_K(q_K^2)$, provided that it is normalized as $f_K(q_K^2=M_K^2)=1$. One sees that, when the interpolated Regge amplitude is employed for $t$-channel $K$ exchange, the replacement of Eq.~(\ref{eq:form_factor_t}) still keeps the the normalization condition of the form factor:
\begin{equation}
\lim_{q^2_K \to M^2_K} {\cal F}_{R,K} = 1.
\end{equation}
Therefore, as soon as we do the same replacement of Eq.~(\ref{eq:form_factor_t}) for the form factor of $t$-channel $K$ exchange everywhere in $C^\mu$ and $M^{\nu\mu}_{\rm int}$, the full photoproduction amplitude still satisfies the generalized Ward-Takahashi identity and, thus, is fully gauge invariant.

\section{Results and discussion}   \label{Sec:results}

As discussed in the introduction section of this paper, the reaction $\gamma p\to K^+\Lambda(1520)$ has been theoretically investigated based on effective Lagrangian approaches by four theory groups in $11$ publications  \cite{Nam:2005uq,Nam:2006cx,Nam:2009cv,Nam:2010au,Toki:2007ab,Xie:2010yk,Xie:2013mua,Wang:2014jxb,He:2012ud,He:2014gga,Yu:2017kng}. The common feature of the results from these theoretical works is that the contributions from the contact term and the $t$-channel $K$ exchange are important for the $\gamma p \to K^+\Lambda(1520)$ reaction. Apart from that, no common ground has been found by these theoretical works for the reaction mechanisms of $\gamma p \to K^+\Lambda(1520)$. In particular, different groups gave quite different answers for the following questions: Are the contributions from the $t$-channel $K^\ast$ exchange and $u$-channel $\Lambda$ exchange significant or not in this reaction, are the nucleon resonances introduced in the $s$ channel indispensable or not to describe the available data, and, if yes, what are the resonance contents and their associated parameters in this reaction? On the other hand, we notice that even though the data on photon-beam asymmetries for $\gamma p \to K^+\Lambda(1520)$ have been reported by the LEPS Collaboration in 2010, they have never been well reproduced in previous theoretical publications of Refs.~\cite{Nam:2005uq,Nam:2006cx,Nam:2009cv,Nam:2010au,Toki:2007ab,Xie:2010yk,Xie:2013mua,Wang:2014jxb,He:2012ud,He:2014gga}. One believes that these photon-beam-asymmetry data will definitely put further constraints on the reaction amplitudes. In Ref.~\cite{Yu:2017kng}, the photon-beam-asymmetry data have indeed been analyzed, but there, the structures of the angular distributions exhibited by the data are missed due to the lack of nucleon resonances. In a word, all previous theoretical publications in regards to $\gamma p\to K^+\Lambda(1520)$ are divided over the reaction mechanism and the resonance contents and parameters of this reaction. A simultaneous description of the differential cross-section data and the photon-beam-asymmetry data still remains to be accomplished.

The purpose of the present work is to get a clear understanding of the reaction mechanism of $\gamma p \to K^+\Lambda(1520)$ based on a combined analysis of the available data on both the differential cross sections and the photon-beam asymmetries within an effective Lagrangian approach. As the differential cross-section data exhibit clear bump structures in the near-threshold region, apart from the $N$, $K$, $K^\ast$, and $\Lambda$ exchanges and the interaction current in the nonresonant background, we introduce as few as possible near-threshold nucleon resonances in the $s$ channel in constructing the $\gamma p \to K^+\Lambda(1520)$ reaction amplitudes to reproduce the data.

\begin{figure}[tbp]
\includegraphics[width=\columnwidth]{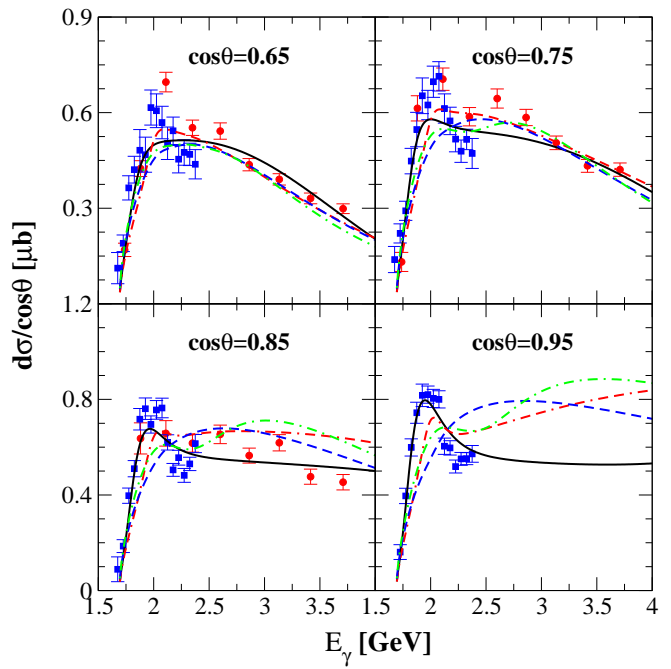}
\caption{Differential cross sections for $\gamma p\to K^+\Lambda(1520)$ at a few selected scattering angles as a function of the photon incident energy. The black solid lines, red dot-double-dashed lines, blue dashed lines, and green dot-dashed lines denote the results obtained by including the $N(2000)5/2^+$, $N(2040)3/2^+$, $N(2100)1/2^+$, and  $N(2190)7/2^-$ resonances in the $s$ channel, respectively. Data are taken from the CLAS Collaboration \cite{Moriya:2013hwg}  (red circles) and the LEPS Collaboration \cite{Kohri:2009xe} (blue squares). For $\cos\theta=0.85$, the CLAS data at $\cos\theta=0.84$ ($E_\gamma < 3.25$ GeV) and $\cos\theta=0.83$ ($E_\gamma > 3.25$ GeV) are shown. }  
\label{fig:dsig-others}
\end{figure}

In the most recent version of RPP \cite{Tanabashi:2018oca}, there are six nucleon resonances near the $K^+\Lambda(1520)$ threshold, namely, the $N(2000)5/2^+$, $N(2040)3/2^+$, $N(2060)5/2^-$, $N(2100)1/2^+$, $N(2120)3/2^-$, and $N(2190)7/2^-$ resonances. If none of these nucleon resonances are introduced in the construction of the $s$-channel reaction amplitudes, we find that it is not possible to achieve a simultaneous description of both the differential cross-section data and the photon-beam-asymmetry data in our model. We then try to reproduce the data by including one of these six near-threshold resonances. If we include one of the $N(2000)5/2^+$, $N(2040)3/2^+$, $N(2100)1/2^+$, and $N(2190)7/2^-$ resonances, we find that the obtained theoretical results for differential cross sections and photon-beam asymmetries have rather poor fitting qualities. As an illustration, we show in Fig.~\ref{fig:dsig-others} the differential cross sections at a few selected scattering angles as a function of the incident photon energy which are obtained by including one of the $N(2000)5/2^+$ (black solid lines), $N(2040)3/2^+$ (red dot-double-dashed lines), $N(2100)1/2^+$ (blue dashed lines), and $N(2190)7/2^-$ (green dot-dashed lines) resonances and compared with the corresponding data \cite{Kohri:2009xe,Moriya:2013hwg}. One sees clearly from Fig.~\ref{fig:dsig-others} that the fits with one of the $N(2040)3/2^+$, $N(2100)1/2^+$, and $N(2190)7/2^-$ resonances fail to describe the differential cross sections at $\cos\theta=0.95$, and the fit with the $N(2000)5/2^+$ resonance fails to reproduce the differential cross-section data at the other three selected scattering angles. In a word, none of these four fits that includes one of the $N(2000)5/2^+$, $N(2040)3/2^+$, $N(2100)1/2^+$, and $N(2190)7/2^-$ resonances can well describe the differential cross-section data. Thus, they are excluded to be acceptable fits. On the other hand, if either the resonance $N(2060)5/2^-$  or the resonance $N(2120)3/2^-$ is considered, a simultaneous description of both the differential cross-section data and the photon-beam-asymmetry data can be satisfactorily obtained, which will be discussed below in detail. Consequently, these two fits, i.e., the ones including the $N(2060)5/2^-$ or the $N(2120)3/2^-$ resonance, are treated as acceptable. When an additional resonance is further included, the fit quality will be improved a little bit, since one has more adjustable model parameters. But, in this case, one would obtain too many solutions with similar fitting qualities, and meanwhile the fitted error bars of adjustable parameters are also relatively large. As a consequence, no conclusive conclusion can be drawn about the resonance contents and parameters extracted from the available data for the considered reaction. We thus conclude that the available differential cross-section data and the photon-beam-asymmetry data for $\gamma p \to K^+\Lambda(1520)$ can be described by including one of the $N(2060)5/2^-$ and $N(2120)3/2^-$ resonances and postpone the analysis of these available data with two or more nucleon resonances until more data for this reaction become available in the future.

\begin{table}[htbp]
\caption{ Fitted values of model parameters. The asterisks below resonance names represent the overall status of these resonances evaluated by RPP \cite{Tanabashi:2018oca}. The numbers in the brackets below the resonance masses and widths denote the corresponding values advocated by RPP \cite{Tanabashi:2018oca}. $\sqrt{\beta_{\Lambda^\ast K}}A_{j}$ represents the reduced helicity amplitude for resonance with $\beta_{\Lambda^\ast K}$ denoting the branching ratio of resonance decay to $\Lambda(1520) K$ and $A_{j}$ standing for the helicity amplitude with spin $j$ for resonance radiative decay to $\gamma p$.  }
\label{table:constants}
\renewcommand{\arraystretch}{1.2}
\begin{tabular*}{\columnwidth}{@{\extracolsep\fill}lcc}
\hline\hline
                                 & Fit A           & Fit B                 \\
\hline
$s_R$ $[{\rm GeV}^2]$       &  $5.17 \pm 0.02$      &  $3.80 \pm 0.12$ 	    \\	
$s_0$ $[{\rm GeV}^2]$       &  $0.81 \pm 0.02$       &  $8.00 \pm 0.05$	    \\
$t_R$ $[{\rm GeV}^2]$       &  $0.80 \pm 0.03$       &  $1.16 \pm 0.07$         \\	
$t_0$ $[{\rm GeV}^2]$        &  $1.60 \pm 0.07$      &   $0.96 \pm 0.07$	   \\
$\Lambda_{B}$ $[{\rm MeV}]$        &  $748 \pm 2$            &   $770 \pm 5$  	     	  \\
$g^{(1)}_{\Lambda^\ast\Lambda\gamma}$              &  $0.00 \pm 0.01$  	&   $8.99 \pm 0.51$   \\
$g_{\Lambda^\ast N K^\ast}$       &   $-22.48 \pm 0.91$    &   $-54.22 \pm 3.72$     \\
\hline
                  &   $N(2060){5/2}^-$   &       $N(2120){3/2}^-$   \\
		                         &  $\ast$$\ast$$\ast$     &  $\ast$$\ast$$\ast$   	\\
$M_R$ $[{\rm MeV}]$                      &  $2020 \pm 1$         &  $2184 \pm 2$        	\\
		                         &  $[2030$$-$$2200]$   &  $[2060$$-$$2160]$   \\
$\Gamma_R$ $[{\rm MeV}]$                 &  $200 \pm 30$    &  $83 \pm 4$          \\
				                 &  $[300$$-$$450]$    &  $[260$$-$$360]$    \\
$\Lambda_{R}$ $[{\rm MeV}]$              &  $1086 \pm 3$        &  $2000 \pm 52$       	\\
$\sqrt{\beta_{\Lambda^\ast K}}A_{1/2}$ $[10^{-3}\,{\rm GeV}^{-1/2}]$  & $3.07 \pm 0.02$   &  $3.04 \pm 0.11$  \\
$\sqrt{\beta_{\Lambda^\ast K}}A_{3/2}$ $[10^{-3}\,{\rm GeV}^{-1/2}]$  & $0.54 \pm 0.02$   &  $5.27 \pm 0.19$  \\
$g^{(2)}_{R\Lambda^\ast K}/g^{(1)}_{R\Lambda^\ast K}$           &  $-1.26 \pm 0.01$   &  $-4.06 \pm 0.28$  \\
\hline\hline
\end{tabular*}
\end{table}

\begin{figure*}[htbp]
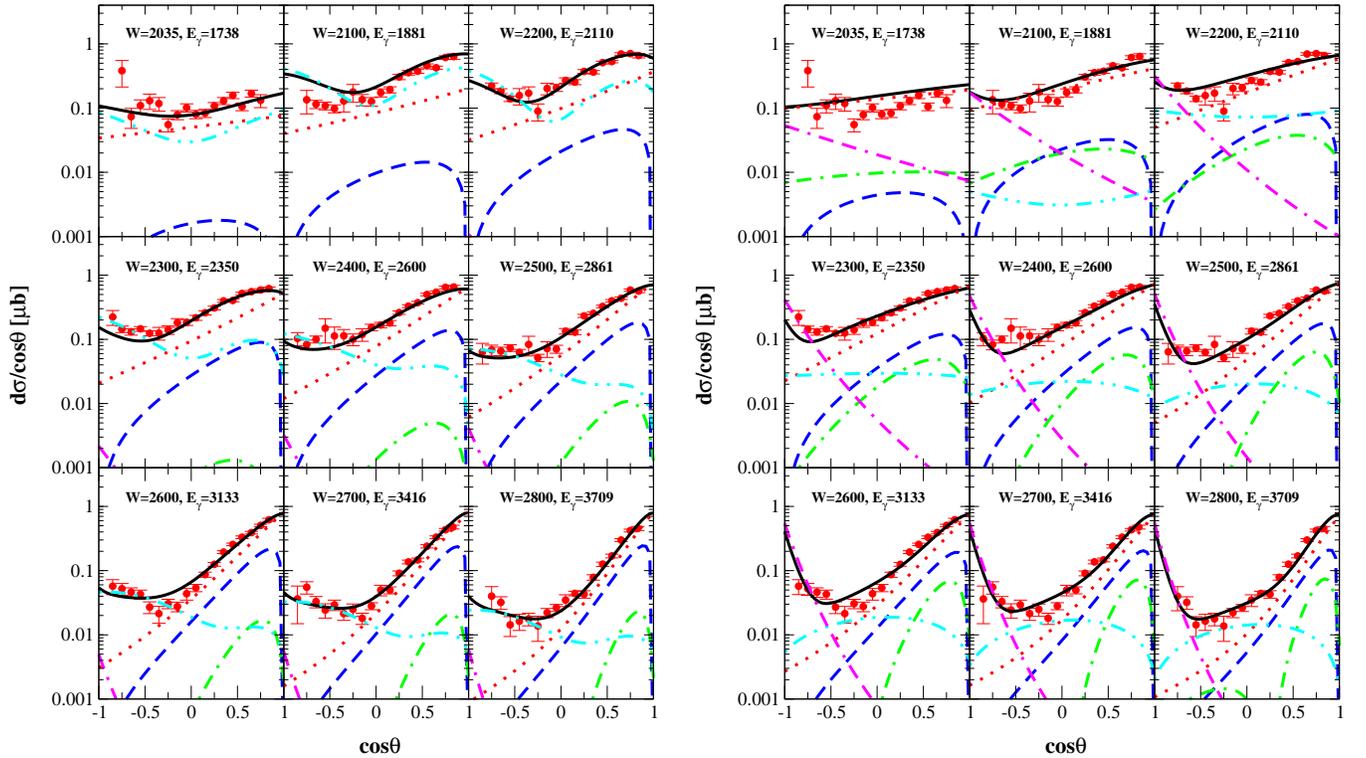

\centering
\includegraphics[width=1.0\columnwidth]{aa_clas_dsigf1.eps}
\hspace{0.1 in}
\includegraphics[width=1.0\columnwidth]{aa_clas_dsigf3.eps}
\caption{Differential cross sections for $\gamma p\to K^+\Lambda(1520)$ as a function of $\cos\theta$ from fit A (left panel) and fit B (right panel). The symbols $W$ and $E_\gamma$ denote the center-of-mass energy of the whole system and the photon laboratory energy, respectively, both in MeV. The black solid lines represent the results calculated from the full amplitudes. The red dotted lines, blue dashed lines, green dot-dashed lines, cyan double-dot-dashed lines, and magenta dot-double-dashed lines denote the individual contributions from the interaction current, the $t$-channel $K$ exchange, the $t$-channel $K^\ast$ exchange, the $s$-channel $N^\ast$ resonance exchange, and the $u$-channel $\Lambda$ exchange, respectively. The scattered symbols are data from the CLAS Collaboration \cite{Moriya:2013hwg}.}
\label{fig:dsig-clas}
\end{figure*}

\begin{figure*}[htbp]
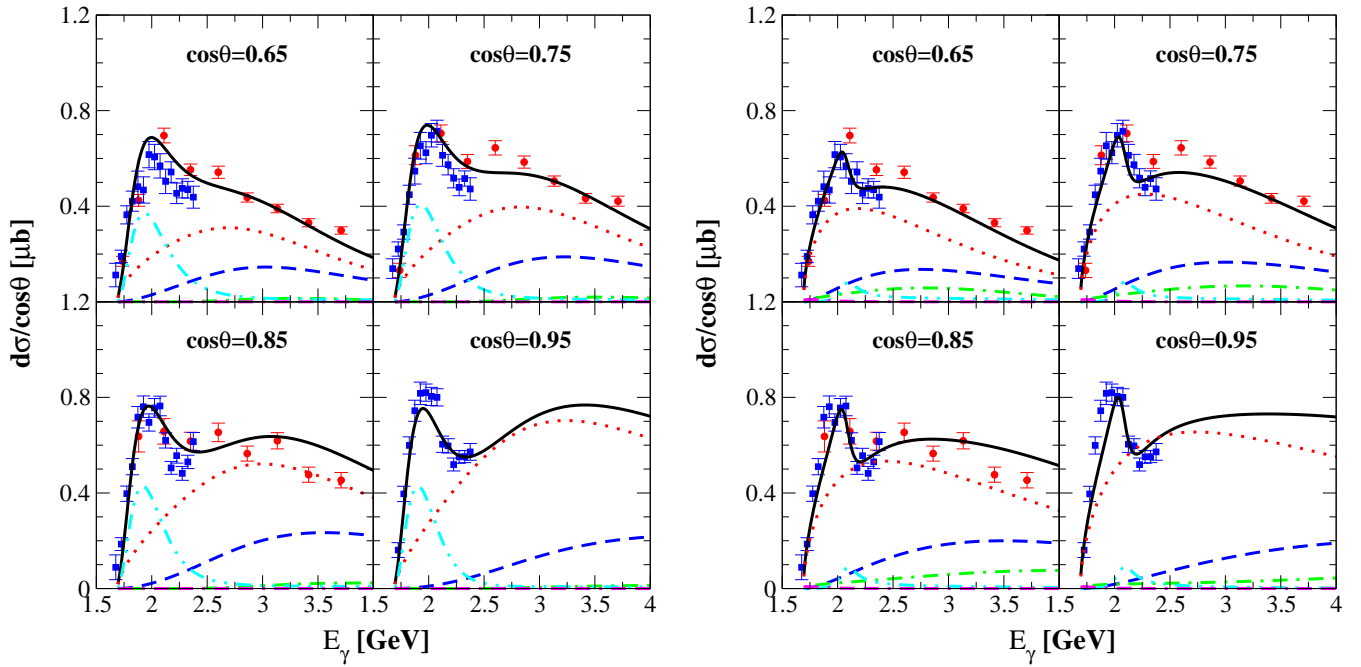

\centering
\includegraphics[width=\columnwidth]{aa_leps_dsigf1.eps}
\hspace{0.1 in}
\includegraphics[width=\columnwidth]{aa_leps_dsigf3.eps}
\caption{Differential cross sections for $\gamma p\to K^+\Lambda(1520)$ at a few selected scattering angles as a function of the photon incident energy from fit A (left panel) and fit B (right panel). The notations for the lines are the same as in Fig.~\ref{fig:dsig-clas}. Data are taken from the CLAS Collaboration \cite{Moriya:2013hwg} (red circles) and the LEPS Collaboration \cite{Kohri:2009xe} (blue squares). For $\cos\theta=0.85$, the CLAS data at $\cos\theta=0.84$ ($E_\gamma < 3.25$ GeV) and $\cos\theta=0.83$ ($E_\gamma > 3.25$ GeV) are shown. }
\label{fig:dsig-leps}
\end{figure*}

\begin{figure}[tbp]
\includegraphics[width=\columnwidth]{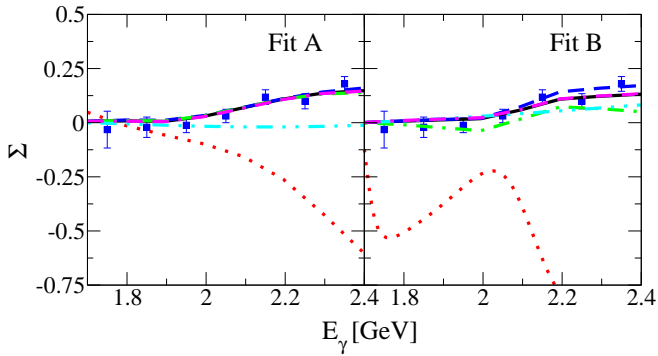}
\caption{Photon-beam asymmetries for $\gamma p\to K^+\Lambda(1520)$ at $\cos\theta=0.8$ as a function of the photon incident energy from fit A (left panel) and fit B (right panel). The black solid lines represent the results calculated from the full amplitudes. The red dotted lines, blue dashed lines, green dot-dashed lines, cyan double-dot-dashed lines, and magenta dot-double-dashed lines denote the results obtained by switching off the contributions of the interaction current, the $t$-channel $K$ exchange, the $t$-channel $K^\ast$ exchange, the $s$-channel $N^\ast$ resonance exchange, and the $u$-channel $\Lambda$ exchange, respectively, from the full model. Data are in the bin $0.6<\cos\theta<1$ and taken from the LEPS Collaboration \cite{Kohri:2009xe}.}
\label{fig:beam-leps}
\end{figure}

As discussed above, we introduce nucleon resonances as few as possible in constructing the reaction amplitudes to describe the available data for $\gamma p \to K^+ \Lambda(1520)$. It is found that a simultaneous description of both the differential cross-section data and the photon-beam-asymmetry data can be achieved by including either the $N(2060)5/2^-$ resonance or the $N(2120)3/2^-$ resonance. We thus get two acceptable fits named as ``fit A," which includes the $N(2060)5/2^-$ resonance, and ``fit B," which includes the $N(2120)3/2^-$ resonance. The fitted values of the adjustable model parameters in these two fits are listed in Table~\ref{table:constants}, and the corresponding results on differential cross sections and photon-beam asymmetries are shown in Figs.~\ref{fig:dsig-clas}$-$\ref{fig:beam-leps}. 

In Table~\ref{table:constants}, for $u$-channel $\Lambda$ exchange, only the value of the coupling constant $g^{(1)}_{\Lambda^\ast\Lambda\gamma}$ is listed. The other coupling constant $g^{(2)}_{\Lambda^\ast\Lambda\gamma}$ is not treated as a free parameter, since it is constrained by the $\Lambda(1520)$ radiative decay width $\Gamma_{\Lambda(1520) \to \Lambda\gamma}=0.133$ MeV as given by RPP \cite{Tanabashi:2018oca}, which results in $g^{(2)}_{\Lambda^\ast\Lambda\gamma}= 2.13$ in fit A and $-13.01$ in fit B, respectively. The asterisks below the resonance names represent the overall status of these resonances evaluated in the most recent RPP \cite{Tanabashi:2018oca}. One sees that both the $N(2060)5/2^-$ and the $N(2120)3/2^-$ resonances are evaluated as three-star resonances. The symbols $M_R$, $\Gamma_R$, and $\Lambda_R$ denote the resonance mass, width, and cutoff parameter, respectively. The numbers in brackets below the resonance mass and width are the corresponding values estimated by RPP. It is seen that the fitted masses of the $N(2060)5/2^-$ and $N(2120)3/2^-$ resonances are comparable with their values quoted by RPP, while the fitted widths for these two resonances are smaller than the corresponding RPP values. For resonance couplings, since in the tree-level calculation only the products of the resonance hadronic and electromagnetic coupling constants are relevant to the reaction amplitudes, we list the reduced helicity amplitudes $\sqrt{\beta_{\Lambda^\ast K}}A_j$ for each resonance instead of showing their hadronic and electromagnetic coupling constants separately \cite{Huang:2013,Wang:2017tpe,Wang:2018vlv,Wang:2019mid}. Here $\beta_{\Lambda^\ast K}$ is the branching ratio for resonance decay to $\Lambda(1520) K$, and $A_j$ is the helicity amplitude with spin $j$ ($j=1/2,3/2$) for resonance radiative decay to $\gamma p$. 

We have, in total, as shown in Figs.~\ref{fig:dsig-clas}$-$\ref{fig:beam-leps}, $220$ data points in the fits. Fit A has a global $\chi^2/N=2.10$, and fit B has a global $\chi^2/N=2.63$. Note that, in the fitting procedure, $11.6\%$ and $5.92\%$ systematic errors for the data from the CLAS Collaboration and the LEPS Collaboration, respectively, have been added in quadrature to the statistical errors \cite{Moriya:2013hwg,Kohri:2009xe}. Overall, one sees that both the differential cross-section data and the photon-beam-asymmetry data have been well described simultaneously in both fit A and fit B.

Figures~\ref{fig:dsig-clas} and \ref{fig:dsig-leps} show the differential cross sections for $\gamma p\to K^+\Lambda(1520)$ resulted from fit A (left panels), which includes the $N(2060)5/2^-$ resonance, and fit B (right panels), which includes the $N(2120)3/2^-$ resonance. There, the black solid lines represent the results calculated from the full reaction amplitudes. The red dotted lines, blue dashed lines, green dot-dashed lines, cyan double-dot-dashed lines, and magenta dot-double-dashed lines denote the individual contributions from the interaction current, the $t$-channel $K$ exchange, the $t$-channel $K^\ast$ exchange, the $s$-channel $N^\ast$ resonance exchange, and the $u$-channel $\Lambda$ exchange, respectively. The individual contributions from the $s$-channel nucleon exchange are too small to be clearly shown in these figures. One sees from Figs.~\ref{fig:dsig-clas} and \ref{fig:dsig-leps} that the differential cross-section data are well reproduced in both fit A (left panels) and fit B (right panels). Note that in Fig.~\ref{fig:dsig-leps}, for $\cos\theta=0.85$, the CLAS data at $\cos\theta=0.84$ ($E_\gamma < 3.25$ GeV) and $\cos\theta=0.83$ ($E_\gamma > 3.25$ GeV) are shown. That explains why in Fig.~\ref{fig:dsig-clas} the theoretical results agree with the CLAS data at high-energy forward angles but in Fig.~\ref{fig:dsig-leps} the theoretical differential cross sections at $\cos\theta=0.85$ overestimate the CLAS data at the last two energy points.

From Figs.~\ref{fig:dsig-clas} and \ref{fig:dsig-leps}, one sees that, in fit A, the contribution from the interaction current [cf. Eq.~(\ref{eq:Mint})] plays a rather important role in the whole energy region. In the near-threshold region, the differential cross sections are dominated by the interaction current and the $N(2060)5/2^-$ resonance exchange. Actually, the contributions from these two terms are responsible for the sharp rise of differential cross sections near the $K^+\Lambda(1520)$ threshold, in particular, the bump structure near $E_\gamma \approx 2$ GeV at forward angles as exhibited by the LEPS data in Fig.~\ref{fig:dsig-leps}. The $t$-channel $K$ exchange is seen to contribute significantly at higher energies and forward angles. The $t$-channel $K^\ast$ exchange has tiny contributions at high-energy forward angles, while the contributions from the $u$-channel $\Lambda$ exchange are negligible. In fit B, the interaction current plays a dominant role in the whole energy region and is also responsible for the sharp rise of the differential cross sections at forward angles near the $K^+\Lambda(1520)$ threshold. The bump structure near $E_\gamma \approx 2$ GeV at forward angles as exhibited by the LEPS data in Fig.~\ref{fig:dsig-leps} is caused by the $N(2120)3/2^-$ resonance on the base of the background dominated by the interaction current. The $t$-channel $K$ exchange and the $u$-channel $\Lambda$ exchange have significant contributions at forward and backward angles, respectively, mostly at higher energies. Considerable contributions are also seen from the $t$-channel $K^\ast$ exchange at high-energy forward angles.

The results of photon-beam asymmetries for $\gamma p\to K^+\Lambda(1520)$ from fit A and fit B are shown, respectively, in the left and right panels in Fig.~\ref{fig:beam-leps}. There, the black solid lines represent the results calculated from the full amplitudes. The red dotted lines, blue dashed lines, green dot-dashed lines, cyan double-dot-dashed lines, and magenta dot-double-dashed lines denote the results obtained by switching off the contributions of the interaction current, the $t$-channel $K$ exchange, the $t$-channel $K^\ast$ exchange, the $s$-channel $N^\ast$ resonance exchange, and the $u$-channel $\Lambda$ exchange, respectively, from the full model. One sees that the photon-beam-asymmetry data are well reproduced in both fits. In fit A, when the contributions of the $N(2060){5/2}^-$ resonance exchange are switched off from the full model, one gets almost zero beam asymmetries. We have checked and found that the $N(2060){5/2}^-$ resonance exchange alone results in negligible beam asymmetries. This means that it is the interference between the $N(2060){5/2}^-$ resonance exchange and the other interaction terms that is crucial for reproducing the experimental values of the beam asymmetries. A similar observation also holds for the interaction current [cf. Eq.~(\ref{eq:Mint})]. The interaction current alone results in almost zero beam asymmetries, but one gets rather negative beam asymmetries when the contributions from the interaction current are switched off from the full model. This means that the interference between the interaction current and the other interaction terms is very important for reproducing the beam asymmetries. Switching off the contributions of the individual terms other than the $N(2060){5/2}^-$ resonance exchange and the interaction current from the full model does not affect too much the theoretical beam asymmetries. In fit B, the interaction current alone is found to result in almost zero beam asymmetries, the same as in fit A. Nevertheless, it is seen from Fig.~\ref{fig:beam-leps} that one gets rather negative beam asymmetries when the contributions of the interaction current are switched off from the full model, showing the importance of the interference of the interaction current and the other interacting terms in photon-beam asymmetries for $\gamma p\to K^+\Lambda(1520)$. Switching off the contributions of the individual terms other than the interaction current from the full model would not affect the theoretical beam asymmetries too much. In Ref.~\cite{Kohri:2009xe}, it is expected that the positive values of the $K^+\Lambda(1520)$ asymmetries indicate a much larger contribution from the $K^\ast$ exchange. In both fit A and fit B of the present work, we have checked and found that the $K^\ast$ exchange alone does result in positive beam asymmetries, but, when the contributions of the $K^\ast$ exchange are switched off from the full model, the calculated beam asymmetries do not change significantly. In particular, the theoretical beam asymmetries are still positive and close to the experimental values when the contributions of the $K^\ast$ exchange are switched off from the full model.

\begin{figure}[tbp]
\includegraphics[width=\columnwidth]{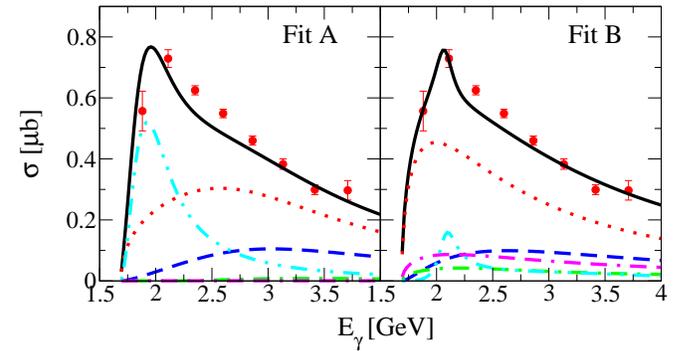}
\caption{Total cross sections for $\gamma p\to K^+\Lambda(1520)$ predicated by fit A (left panel) and fit B (right panel). Notations for the lines are the same as in Fig.~\ref{fig:dsig-clas}. Data are taken from the CLAS Collaboration \cite{Moriya:2013hwg} but not included in the fits.}
\label{fig:total}
\end{figure}

Figure~\ref{fig:total} shows the total cross sections for $\gamma p\to K^+\Lambda(1520)$ predicated from fit A (left panel) and fit B (right panel), which are obtained by integrating the corresponding differential cross sections calculated in these two fits. In Fig.~\ref{fig:total}, the black solid lines represent the results calculated from the full reaction amplitudes. The red dotted lines, blue dashed lines, green dot-dashed lines, cyan double-dot-dashed lines, and magenta dot-double-dashed lines denote the individual contributions from the interaction current, the $t$-channel $K$ exchange, the $t$-channel $K^\ast$ exchange, the $s$-channel $N^\ast$ resonance exchange, and the $u$-channel $\Lambda$ exchange, respectively. The individual contributions from the $s$-channel nucleon exchange are too small to be clearly shown in these figures. Note that the data for the total cross sections of $\gamma p\to K^+\Lambda(1520)$ are not included in the fits. Even so, one sees that, in both fit A and fit B, the theoretical total cross sections are in good agreement with the data. In fit A, the $s$-channel $N(2060){5/2}^-$ exchange, the interaction current, and the $t$-channel $K$ exchange provide the most important contributions to the total cross sections, while the contributions from the $u$-channel $\Lambda$ exchange, the $s$-channel $N$ exchange, and the $t$-channel $K^\ast$ exchange are negligible. The bump structure near $E_\gamma\approx 2$ GeV is caused mainly by the $N(2060){5/2}^-$ resonance exchange and the interaction current. The sharp rise of the total cross sections near the $K^+\Lambda(1520)$ threshold is dominated by the $s$-channel $N(2060){5/2}^-$ exchange. In fit B, the dominant contributions to the total cross sections come from the interaction current, which is also responsible for the sharp rise of the total cross sections near the $K^+\Lambda(1520)$ threshold. The individual contributions from the $s$-channel $N(2120){3/2}^-$ exchange, the $t$-channel $K$ and $K^\ast$ exchanges, and the $u$-channel $\Lambda$ exchange are considerable, while those from the $s$-channel $N$ exchange are negligible to the total cross sections. Comparing the individual contributions in fit A and fit B, one sees that the contributions from the resonance exchange are rather important in fit A, but they are much smaller in fit B. The contributions from the $t$-channel $K^\ast$ exchange and the $u$-channel $\Lambda$ exchange are negligible in fit A, but they are considerable in fit B. In both fits, the interaction current provides dominant contributions, and the $t$-channel $K$ exchange results in considerable contributions to the cross sections.

As mentioned in the introduction section, the $K^+\Lambda(1520)$ photoproduction reaction has been theoretically investigated based on effective Lagrangian approaches by four theory groups in $11$ publications  \cite{Nam:2005uq,Nam:2006cx,Nam:2009cv,Nam:2010au,Toki:2007ab,Xie:2010yk,Xie:2013mua,Wang:2014jxb,He:2012ud,He:2014gga,Yu:2017kng}. In these previous publications, the photon-beam-asymmetry data reported by the LEPS Collaboration in 2010 \cite{Kohri:2009xe} have never been well reproduced except in Ref.~\cite{Yu:2017kng}. But in Ref.~\cite{Yu:2017kng}, the structures of the angular distributions exhibited by the data are missed due to the lack of nucleon resonances in the employed Reggeized model. As shown in Figs.~\ref{fig:dsig-clas}$-$\ref{fig:beam-leps}, the present work for the first time presents a simultaneous description of the data on both differential cross sections and photon-beam asymmetries within an effective Lagrangian approach. The common feature of the results from the previous works \cite{Nam:2005uq,Nam:2006cx,Nam:2009cv,Nam:2010au,Toki:2007ab,Xie:2010yk,Xie:2013mua,Wang:2014jxb,He:2012ud,He:2014gga,Yu:2017kng} is that the contributions from the contact term and the $t$-channel $K$ exchange are important for the $\gamma p \to K^+\Lambda(1520)$ reaction. This feature has also been observed in the present work, as illustrated in Fig.~\ref{fig:total}. The contributions of nucleon resonance exchanges are reported to be small in Refs.~\cite{Nam:2005uq,Nam:2006cx,Nam:2009cv,Nam:2010au}, while the $N(2120)3/2^-$ exchange is found to be important to the cross sections of $\gamma p\to K^+\Lambda(1520)$ in Refs.~\cite{Toki:2007ab,Xie:2010yk,Xie:2013mua,Wang:2014jxb,He:2012ud,He:2014gga}. In the present work, we found that, to get a satisfactory description of the data on both differential cross sections and photon-beam asymmetries of $\gamma p\to K^+\Lambda(1520)$, the exchange of at least one nucleon resonance in the $s$ channel needs to be introduced in constructing the reaction amplitudes. The required nucleon resonance could be either the $N(2060){5/2}^-$ or the $N(2120){3/2}^-$, both evaluated as three-star resonances in the most recent version of RPP \cite{Tanabashi:2018oca}. In the fit with the $N(2060){5/2}^-$ resonance, the contributions of the resonance exchange are found to be rather important to the cross sections, and, in particular, they are responsible for the sharp rise of the cross sections near the $K^+\Lambda(1520)$ threshold, as can be seen in Fig.~\ref{fig:total}. In the fit with the $N(2120){3/2}^-$ resonance, although much smaller than those of the interaction current, the contributions of the resonance exchange are still considerable to the cross sections. In Refs.~\cite{Nam:2005uq,Nam:2006cx,Nam:2009cv,Nam:2010au,Toki:2007ab,Xie:2010yk,Xie:2013mua,Wang:2014jxb,Yu:2017kng}, the $t$-channel $K^\ast$ exchange is found to provide negligible contributions. In Refs.~\cite{He:2012ud,He:2014gga}, it is reported that the contributions of the $t$-channel $K^\ast$ exchange are considerable to the cross sections. In our present work, the contributions of the $t$-channel $K^\ast$ exchange are negligible in the fit with the $N(2060){5/2}^-$ resonance, and are considerable in the fit with the $N(2120){3/2}^-$ resonance. As for the $u$-channel $\Lambda$ exchange, important contributions are reported in Refs.~\cite{Toki:2007ab,Xie:2010yk,Xie:2013mua,Wang:2014jxb,He:2012ud,He:2014gga}, while in the present work, considerable contributions of this term are seen only in the fit with the $N(2120){3/2}^-$ resonance.

\begin{figure}[tbp]
\includegraphics[width=\columnwidth]{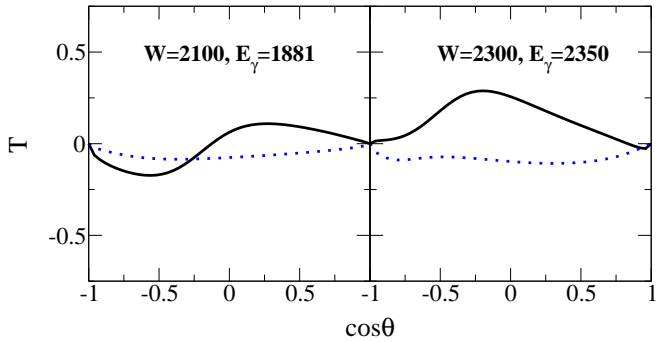}
\caption{Predictions of target nucleon asymmetries for $\gamma p \to K^+\Lambda(1520)$ from fit A (black solid lines) and fit B (blue dashed lines) at two selected center-of-mass energies.}
\label{fig:T}
\end{figure}

From Figs.~\ref{fig:dsig-clas}$-$\ref{fig:total} one sees that the fit with the $N(2060){5/2}^-$ resonance (fit A) and the fit with the $N(2120){3/2}^-$ resonance (fit B) describe the data on differential cross sections and photon-beam asymmetries for $\gamma p\to K^+\Lambda(1520)$ almost equally well. In Fig.~\ref{fig:T}, we show the predictions of the target nucleon asymmetries ($T$) from fit A (black solid lines) and fit B (blue dashed lines) at two selected center-of-mass energies. One sees that unlike the differential cross sections and the photon-beam asymmetries, the target nucleon asymmetries predicted by fit A and fit B are quite different. Future experimental data on target nucleon asymmetries are expected to be able to distinguish the fit A and fit B of the present work and to further clarify the resonance content, the resonance parameters, and the reaction mechanism for the $\gamma p\to K^+\Lambda(1520)$ reaction.

\section{Summary and conclusion}  \label{sec:summary}

The photoproduction reaction $\gamma p\to K^+\Lambda(1520)$ is of interest since the $K^+\Lambda(1520)$ has isospin $1/2$, excluding the contributions of the $\Delta$ resonances from the reaction mechanisms, and the threshold of $K^+\Lambda(1520)$ is at $2.01$ GeV, making this reaction more suitable than $\pi$ production reactions to study the nucleon resonances in a less-explored higher resonance mass region.

Experimentally, the data for $\gamma p\to K^+\Lambda(1520)$ on differential cross sections, total cross sections, and photon-beam asymmetries are available from several experimental groups \cite{Boyarski:1970yc,Barber:1980zv,Kohri:2009xe,Wieland:2010cq,Moriya:2013hwg}, with the photon-beam-asymmetry data coming from the LEPS Collaboration \cite{Kohri:2009xe} and the most recent differential and total cross-section data coming from the CLAS Collaboration \cite{Moriya:2013hwg}.

Theoretically, the cross-section data for $\gamma p\to K^+\Lambda(1520)$ have been analyzed by several theoretical groups \cite{Nam:2005uq,Nam:2006cx,Nam:2009cv,Nam:2010au,Toki:2007ab,Xie:2010yk,Xie:2013mua,Wang:2014jxb,He:2012ud,He:2014gga} within effective Lagrangian approaches, and the photon-beam-asymmetry data \cite{Kohri:2009xe} have been reproduced only in Ref.~\cite{Yu:2017kng} within a Reggeized framework. In the latter, the apparent structures of the angular distributions exhibited by the data are missing due to the lack of nucleon resonances in $s$-channel interactions in the Regge model. In these publications, the reported common feature for the $\gamma p\to K^+\Lambda(1520)$ reaction is that the contributions from the contact term and the $t$-channel $K$ exchange are important to the cross sections of this reaction. Nevertheless, the reaction mechanisms of $\gamma p \to K^+\Lambda(1520)$ claimed by different theoretical groups are quite different. In particular, there are no conclusive answers for the questions of whether the contributions from the $t$-channel $K^\ast$ exchange and $u$-channel $\Lambda$ exchange are significant or not, whether the introduction of nucleon resonances in the $s$ channel is inevitable or not for describing the data, and if yes, what resonance contents and parameters are needed in this reaction.

In the present work, we performed a combined analysis of the data on both the differential cross sections and photon-beam asymmetries for $\gamma p \to K^+\Lambda(1520)$ within an effective Lagrangian approach. We considered the $t$-channel $K$ and $K^\ast$ exchange, the $u$-channel $\Lambda$ exchange, the $s$-channel nucleon and nucleon resonance exchanges, and the interaction current, with the last one being constructed in such a way that the full photoproduction amplitudes satisfy the generalized Ward-Takahashi identity and, thus, are fully gauge invariant. The strategy for introducing the nucleon resonances in the $s$ channel used in the present work was that we introduce nucleon resonances as few as possible to describe the data. 

For the first time, we achieved a satisfactory description of the data on both the differential cross sections and the photon-beam asymmetries for $\gamma p \to K^+\Lambda(1520)$. We found that either the $N(2060){5/2}^-$ or the $N(2120){3/2}^-$ resonance needs to be introduced in constructing the $s$-channel reaction amplitudes in order to get a simultaneous description of the data on differential cross sections and photon-beam asymmetries for $\gamma p \to K^+\Lambda(1520)$. In both cases, the contributions of the interaction current and the $t$-channel $K$ exchange are found to dominate the background contributions. The $s$-channel resonance exchange is found to be rather important in the fit with the $N(2060){5/2}^-$ resonance and to be much smaller but still considerable in the fit with the $N(2120){3/2}^-$ resonance. The contributions of the $t$-channel $K^\ast$ exchange and the $u$-channel $\Lambda$ exchange are negligible in the fit with the $N(2060){5/2}^-$ resonance and are significant in the fit with the $N(2120){3/2}^-$ resonance. The target nucleon asymmetries for $\gamma p \to K^+\Lambda(1520)$ are predicted, on which the future experimental data are expected to verify our theoretical models, to distinguish the two fits with either the $N(2060){5/2}^-$ or the $N(2120){3/2}^-$ resonance, and to further clarify the reaction mechanisms of the $K^+\Lambda(1520)$ photoproduction reaction.

\begin{acknowledgments}
This work is partially supported by the National Natural Science Foundation of China under Grants No.~11475181 and No.~11635009, the Fundamental Research Funds for the Central Universities, and the Key Research Program of Frontier Sciences of the Chinese Academy of Sciences under Grant No.~Y7292610K1.
\end{acknowledgments}

\end{document}